\documentclass[conference]{IEEEtran}
\IEEEoverridecommandlockouts
\usepackage[T1]{fontenc}
\usepackage{cite}
\usepackage{amsmath,amssymb,amsfonts}
\usepackage{graphicx,booktabs,array}
\usepackage{algorithm,algpseudocode}
\usepackage{flushend}
\usepackage[hidelinks]{hyperref}

\graphicspath{{figures/}}
\usepackage{xcolor}

\begin{document}
\setlength{\dbltextfloatsep}{5pt plus 1pt minus 2pt}
\setlength{\dblfloatsep}{3pt plus 1pt minus 1pt}

\title{Electric Vehicle Scheduling and Carbon-Aware Energy Dispatch for Charging Stations%
\thanks{\scriptsize
This work was supported in part by the Clean Energy and Equitable
Transportation Solutions (CLEETS) NSF--UKRI Global Centre under NSF
Award No.\ 2330565 and UKRI Award No.\ EP/Y026233/1.
\textit{(*Corresponding author: Fulong Yao.)}
}
}

\author{
\IEEEauthorblockN{\normalsize
Fulong Yao\textsuperscript{1,*},
Yiming Xu\textsuperscript{2},
Maurizio Albano\textsuperscript{2},
Liana Cipcigan\textsuperscript{2},
Naeima Hamed\textsuperscript{1},
Shangyin Weng\textsuperscript{1},
Omer Rana\textsuperscript{1}
}
\IEEEauthorblockA{\small
\textsuperscript{1}School of Computational and Mathematical Sciences, Cardiff University, Cardiff, UK\\
\textsuperscript{2}School of Engineering, Cardiff University, Cardiff, UK\\
\{yaof, XuY130, AlbanoM, CipciganLM, HamedNH, WengS3, RanaOF\}@cardiff.ac.uk
}
}
\maketitle

\begin{abstract}
Electric vehicle (EV) charging stations integrating photovoltaic (PV) generation, battery storage, and bidirectional grid exchange must coordinate access to limited chargers with energy supply while considering both energy cost and carbon emissions. This paper proposes a solver-free greedy heuristic (SFGH) for dynamic vehicle scheduling (VS) and a twin delayed deep deterministic policy gradient (TD3) controller for carbon-aware energy dispatch (ED). SFGH assigns EVs according to reductions in service delays, and TD3 controls battery operation to balance energy costs and carbon emissions while meeting the resulting charging demand. The model allows both PV and battery exports, accounting for export revenue and carbon credits. Simulations on year-long case studies show that SFGH--TD3 achieves the lowest accumulated service delay, energy cost, and carbon emissions among the evaluated heuristic and multi-agent deep reinforcement learning (DRL) configurations.

\end{abstract}


\section{Introduction}\label{sec:1}
Electric vehicle (EV) charging stations provide essential  infrastructure needed to support transport electrification. Their operation connects EV charging requirements with energy supply, making effective management important for both service quality and energy use~\cite{he2025coordinated}. Integrating photovoltaic (PV) generation and energy storage systems (ESS) creates opportunities to supply vehicles with local renewable energy and shift grid purchases across time. Realizing these benefits requires coordinating charging service with available energy resources~\cite{liu2025hierarchical}. This coordination involves two closely related tasks: vehicle scheduling (VS), which determines how waiting vehicles access limited chargers; and energy dispatch (ED), which controls PV, storage, and grid exchange to supply the resulting charging demand.

Differences in vehicle arrival times and per-vehicle energy requirements complicate VS, while variable PV output and time-varying electricity prices require ED to adapt its energy allocation decisions~\cite{fan2024agent}. Bidirectional grid exchange and battery storage provide opportunities to reduce net energy costs through surplus PV sales and energy arbitrage~\cite{yao2025unified}. Beyond these economic benefits, the decarbonization goals of transport electrification also call for explicit consideration of charging-related carbon emissions. Since grid carbon intensity varies with the generation mix, minimizing electricity expenditure alone does not necessarily minimize these emissions~\cite{yao2025holistic}. Effective coordination between VS and ED therefore requires maintaining charging service while accounting for both the financial and carbon consequences of energy decisions.

Cost-oriented studies seek to reduce charging-station operational costs while meeting vehicle service requirements. Wu et al.~\cite{wu2024dynamic} developed a dynamic energy management system that used charging demand, PV generation, and battery status to schedule storage operation, reduce electricity costs, and improve station profitability. Fan et al.~\cite{fan2024agent} employed charger-level multi-agent deep reinforcement learning (DRL) to control PV-assisted charging and vehicle-to-grid operation, reducing energy costs and improving charging satisfaction under uncertain conditions and charger faults. Addressing the connection between VS and ED explicitly, Yao et al.~\cite{yao2026hierarchical} adopted heuristic VS to endogenously determine charging and refuelling demand profiles, which were then
passed to DRL-based ED in an integrated electric--hydrogen--transport system. These studies provide approaches to economic operation and service coordination, but financial performance alone does not establish the carbon benefits of dispatch decisions.

Carbon-aware studies explicitly incorporate the environmental consequences of electricity use. Cheng et al.~\cite{cheng2022carbon} combined carbon-intensity forecasting with rolling-horizon charging optimization to balance emissions reduction and charging energy delivery under user schedules and infrastructure constraints. Considering both economic and environmental objectives, Li et al.~\cite{li2026tariff} developed a rule-based supervisory energy management system, where tariff and carbon-intensity signals were used to optimise a PV--battery EV charging station. However, these studies provide limited insight into how VS can be combined with carbon-aware ED to manage access to limited chargers and balance electricity costs and carbon emissions.

To address this challenge, this paper presents a sequential approach for VS and carbon-aware ED. This paper makes two contributions. First, it formulates station operation with explicit charger occupancy, PV and battery exports, and separate financial and carbon accounts. Second, it combines a solver-free greedy heuristic (SFGH) for vehicle scheduling with a twin delayed deep deterministic policy gradient (TD3)~\cite{fujimoto2018addressing} for feasible energy dispatch. SFGH selects vehicle--charger assignments according to reductions in service delays, while TD3 balances financial and carbon objectives to meet the charging demand. 

\section{System Modelling and Problem Formulation} \label{sec:2}
\begin{figure}[t]
\centering
\includegraphics[width=7.0cm, trim=18 18 18 18, clip]{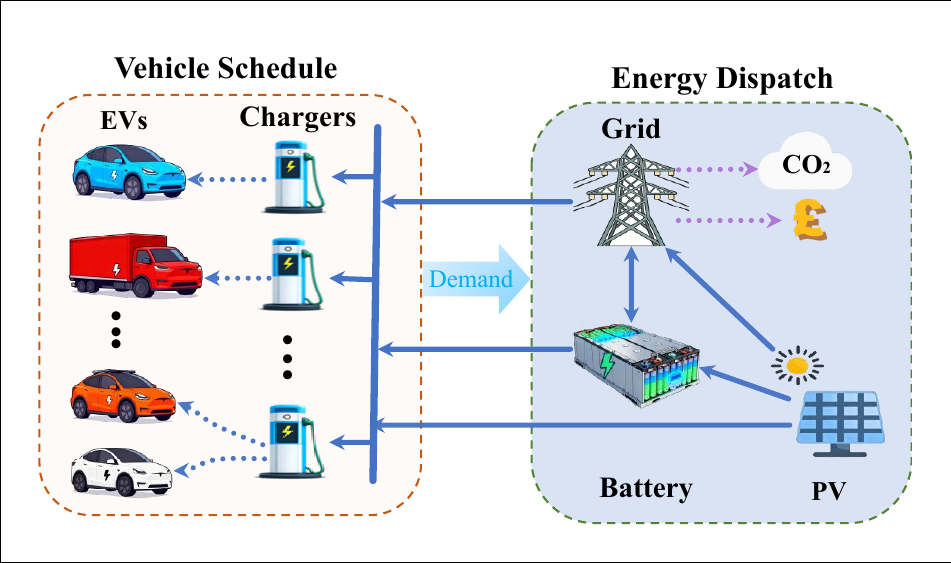}
\caption{Architecture of the EV charging station.}
\label{fig:1}
\end{figure}
Fig.~\ref{fig:1} shows the station architecture. The VS side comprises a set of waiting EVs in a local waiting pool and $N^{ec}$ identical chargers. VS assigns waiting EVs to available chargers and passes the resulting aggregate charging demand to ED. EVs only receive electricity and cannot discharge to the station or the grid. The ED side includes local PV generation, a battery, and a bidirectional grid connection. PV can supply EV demand, charge the battery, or export to the grid. The battery can store electricity from both PV and the grid for later use in EV charging or export to the grid, enabling energy arbitrage over time. Given the charging demand, ED controls battery operation based on electricity prices and grid carbon intensity. Time is divided into intervals of $\Delta t$ hours, with power expressed as interval averages in kW and energy measured in kWh.

\subsection{Vehicle Scheduling}  \label{sec:2-1}
Let $\mathcal S_t^{ev}$ denote the set of EVs waiting at the beginning of interval $t$. Each vehicle $i$ is characterized by its remaining energy demand $E_{i,t}^{ev}$ and accumulated waiting time $w_{i,t}^{ev}$. Each charger serves at most one EV at a time, and each EV uses at most one charger. Once charging begins, it continues without interruption until the vehicle's energy demand is met, potentially spanning multiple intervals. If charging finishes before the end of an interval, the released charger can serve another waiting EV for the remainder of that interval.

Each charger operates at a fixed rated power $P^{ev,rt}$ while charging an EV, with charging efficiency $\eta^{ev}$. Let $\mathcal S_{p,t}^{ev}$ denote the set of EVs served by charger $p$ during interval $t$, including vehicles whose charging began in an earlier interval. If $\tau_{i,p,t}^{ev}$ is the charging duration of vehicle $i$ on charger $p$ within this interval, the total occupied time is
\begin{equation}
\small
\tau_{p,t}^{ev}
=\sum_{i\in\mathcal S_{p,t}^{ev}}\tau_{i,p,t}^{ev},
\qquad 0<\tau_{i,p,t}^{ev}\leq\tau_{p,t}^{ev}\leq\Delta t.
\label{eq:1}
\end{equation}
If $\mathcal S_{p,t}^{ev}=\varnothing$, then $\tau_{p,t}^{ev}=0$. The energy delivered to these EVs is therefore $\eta^{ev}P^{ev,rt}\tau_{p,t}^{ev}$ and the aggregate power demand is
\begin{equation}
\small
P_t^{ev}
=\frac{P^{ev,rt}}{\Delta t}
\sum_{p=1}^{N^{ec}}\tau_{p,t}^{ev}.
\label{eq:2}
\end{equation}

For each waiting EV $i\in\mathcal S_t^{ev}$, let $\delta_{i,t}$
denote its additional waiting time during interval $t$.
It equals the time from the beginning of the interval until
charging starts, or $\Delta t$ if charging does not start
within the interval. The service objective accounts for both waiting time and overflow time (waiting beyond the threshold $w^{max}$) and is expressed as
\begin{equation}
\small
C_t^{wt}=\sum_{i\in\mathcal S_t^{ev}}\delta_{i,t},
\label{eq:3}
\end{equation}
\begin{equation}
\small
C_t^{ot}=\sum_{i\in\mathcal S_t^{ev}}\Bigl(
[w_{i,t}^{ev}+\delta_{i,t}-w^{max}]_+
-[w_{i,t}^{ev}-w^{max}]_+\Bigr),
\label{eq:4}
\end{equation}
where $[x]_+=\max(x,0)$. The waiting cost $C_t^{wt}$ measures total waiting accrued during interval $t$, while the overflow cost $C_t^{ot}$ measures the portion accrued beyond $w^{max}$.
Both are measured in vehicle-hours and include all waiting EVs.

\subsection{Energy Dispatch} \label{sec:2-2}
The battery has energy capacity $E^{bt,cp}$, power rating $P^{bt,max}$, and charging/discharging efficiencies $\eta^{bt,ch}$/$\eta^{bt,dc}$. Its charging and discharging powers, $P_t^{bt,ch}$ and $P_t^{bt,dc}$, satisfy
\begin{equation}
\small
0\leq P_t^{bt,ch},\,P_t^{bt,dc}\leq P^{bt,max},
\label{eq:5}
\end{equation}
\begin{equation}
\small
P_t^{bt,ch}P_t^{bt,dc}=0.
\label{eq:6}
\end{equation}
The state of charge (SOC) evolves as
\begin{equation}
\small
SOC_{t+1}^{bt}
=SOC_t^{bt}+\frac{\Delta t}{E^{bt,cp}}
\left(
\eta^{bt,ch}P_t^{bt,ch}
-\frac{P_t^{bt,dc}}{\eta^{bt,dc}}
\right).
\label{eq:7}
\end{equation}
To maintain SOC within the allowable range $[SOC^{bt,min},SOC^{bt,max}]$, the charging and discharging powers must also meet
\begin{equation}
\small
\left\{
\begin{aligned}
P_t^{bt,ch}
&\leq
\frac{(SOC^{bt,max}-SOC_t^{bt})E^{bt,cp}}
{\eta^{bt,ch}\Delta t},\\
P_t^{bt,dc}
&\leq
\frac{(SOC_t^{bt}-SOC^{bt,min})E^{bt,cp}\eta^{bt,dc}}
{\Delta t}.
\end{aligned}
\right.
\label{eq:8}
\end{equation}

Let $P_t^{pv}$ denote the available PV power, which is always prioritized to meet EV charging demand. The resulting net demand is given by $P_t^{nd}=P_t^{ev}-P_t^{pv}$, where a positive value indicates demand remaining after PV supply, while a negative value indicates surplus PV available for battery charging or grid export. The net grid import $P_t^{gd}$ is then determined by 
\begin{equation}
\small
P_t^{gd}=P_t^{nd}+P_t^{bt,ch}-P_t^{bt,dc}.
\label{eq:10}
\end{equation}
Grid imports and exports are $P_t^{gd,in}=[P_t^{gd}]_+$ and $P_t^{gd,out}=[-P_t^{gd}]_+$, respectively, and are mutually exclusive. 

Let $p_t^{p}$ and $p_t^{s}$ be the grid purchase and sale prices in GBP/kWh, and $c_t$ the grid carbon intensity in gCO$_2$/kWh. The net electricity cost and grid-related carbon emissions are
\begin{align}
\small
C_t^{gd}&=\Delta t\left(P_t^{gd,in}p_t^{p}-P_t^{gd,out}p_t^{s}\right),
\label{eq:11}\\
C_t^{ce}&=\frac{\Delta t}{1000}
\left(P_t^{gd,in}-P_t^{gd,out}\right)c_t.
\label{eq:12}
\end{align}
Exports receive a carbon credit at the same intensity $c_t$ used for imports. $C_t^{ce}$ therefore represents net carbon emissions under this accounting convention. The financial and carbon terms are measured in GBP and kgCO$_2$, respectively.
Over the operating horizon, VS aims to reduce accumulated waiting and overflow delays, while ED balances net electricity cost and net grid-related emissions under the charging demand caused by VS.

\section{Approach} \label{sec:3}
\begin{figure}[!t]
\centering
\includegraphics[width=9.0cm, trim=18 23 18 19, clip]{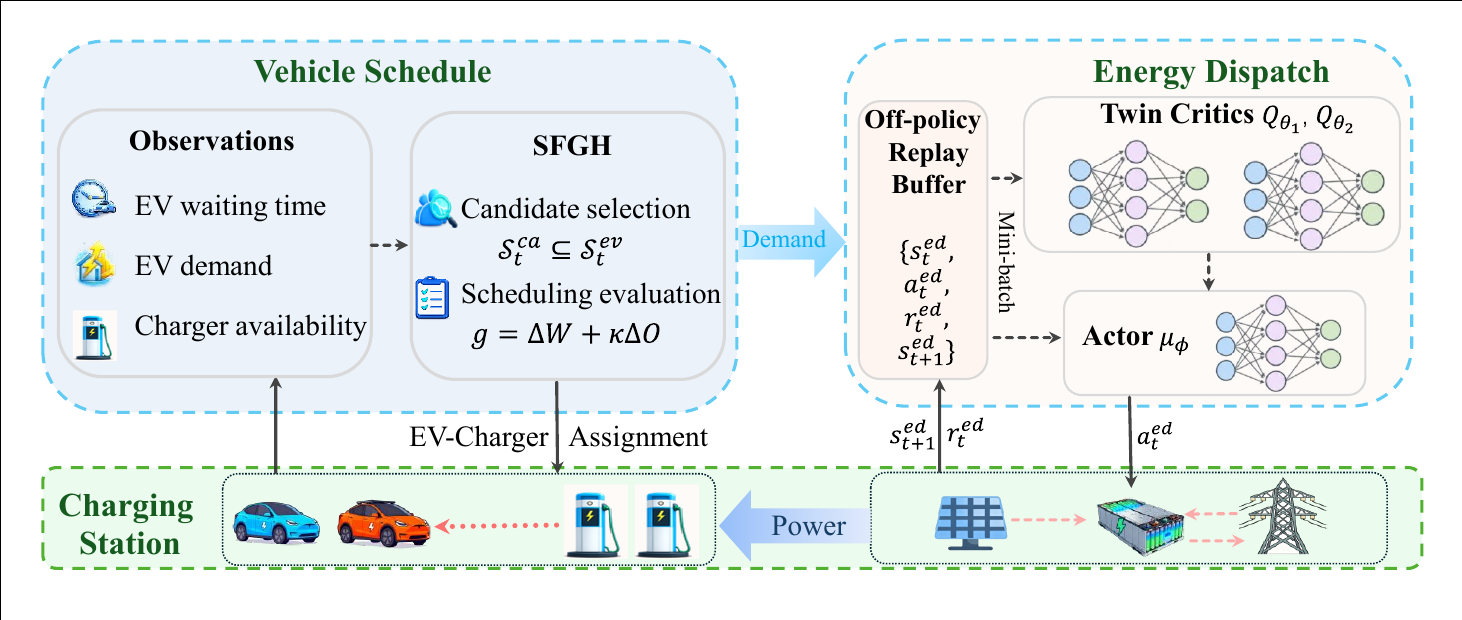}
\caption{The proposed VS-ED control approach}
\label{fig:2}
\end{figure}

Fig.~\ref{fig:2} illustrates the proposed VS--ED control approach, which combines a solver-free greedy heuristic (SFGH) for VS with a twin delayed deep deterministic policy gradient (TD3)-based controller for ED. At each interval, SFGH uses observations of EV waiting times, energy demands, and charger availability to determine vehicle--charger assignments and the resulting charging demand. Given this demand and the available PV generation, TD3 controls battery charging and discharging based on electricity prices and grid carbon intensity. The station then advances to the next interval and provides updated observations.

\subsection{SFGH for Vehicle Scheduling} \label{sec:3-1}
SFGH constructs the schedule through candidate selection and iterative assignment evaluation. It first selects at most $M$ vehicles from the waiting set, forming a candidate set $\mathcal S_t^{ca}\subseteq\mathcal S_t^{ev}$. Vehicles are selected in order of decreasing overflow $[w_{i,t}^{ev}-w^{max}]_+$, with ties broken by longer waiting time and then smaller remaining energy demand. This prioritizes delayed EVs while limiting the number of assignments evaluated, thereby reducing the computational cost of scheduling.

Let $\tau_{p,t}^{av}$ denote the time from the beginning of interval $t$ until charger $p$ becomes available. 
A candidate $i\in\mathcal S_t^{ca}$ can start charging on charger $p$ within the interval if $\tau_{p,t}^{av}<\Delta t$. For each feasible assignment, SFGH compares two cases: EV $i$ remains waiting throughout the interval, giving $\delta_{i,t}=\Delta t$; or starts charging at $\tau_{p,t}^{av}$, giving $\delta_{i,t}=\tau_{p,t}^{av}$. 
Using the costs defined in \eqref{eq:3} and \eqref{eq:4}, the corresponding reductions are
\begin{equation}
\small
\Delta W_{i,p,t}
=\left.C_t^{wt}\right|_{\delta_{i,t}=\Delta t}
-\left.C_t^{wt}\right|_{\delta_{i,t}=\tau_{p,t}^{av}}
=\Delta t-\tau_{p,t}^{av},
\label{eq:13}
\end{equation}
\begin{equation}
\small
\begin{aligned}
\Delta O_{i,p,t}
&=\left.C_t^{ot}\right|_{\delta_{i,t}=\Delta t}
-\left.C_t^{ot}\right|_{\delta_{i,t}=\tau_{p,t}^{av}}\\
&=[w_{i,t}^{ev}+\Delta t-w^{max}]_+ -[w_{i,t}^{ev}+\tau_{p,t}^{av}-w^{max}]_+.
\end{aligned}
\label{eq:14}
\end{equation}
The scheduling evaluation in Fig.~\ref{fig:2} combines these reductions
\begin{equation}
\small
g_{i,p,t}
=\Delta W_{i,p,t}+\kappa\Delta O_{i,p,t},
\label{eq:15}
\end{equation}
where $\kappa\geq0$ weights the overflow reduction. The score measures the decrease in $C_t^{wt}+\kappa C_t^{ot}$ achieved by the  assignment.

SFGH selects the feasible pair with the largest score. The selected EV waits for $\delta_{i,t}=\tau_{p,t}^{av}$, and its charging duration within the interval is
\begin{equation}
\small
\tau_{i,p,t}^{ev}
=\min\left\{
\frac{E_{i,t}^{ev}}{\eta^{ev}P^{ev,rt}},
\,\Delta t-\tau_{p,t}^{av}
\right\}.
\label{eq:16}
\end{equation}
The charger's next available time is updated as $\tau_{p,t}^{av}\leftarrow\tau_{p,t}^{av} +E_{i,t}^{ev}/(\eta^{ev}P^{ev,rt})$, using the full charging duration required by the selected EV. If charging extends beyond the interval, the EV remains connected until its energy demand is met. SFGH removes the assigned EV from $\mathcal S_t^{ca}$ and re-evaluates the remaining feasible pairs. This process repeats until the candidate set is empty or no charger becomes available within the interval. Existing and newly assigned charging sessions determine $P_t^{ev}$ through \eqref{eq:1}--\eqref{eq:2}, which is passed to ED before execution. The waiting and overflow delays in \eqref{eq:3}--\eqref{eq:4} are evaluated over all EVs in $\mathcal S_t^{ev}$, including those excluded during candidate selection.

\subsection{TD3 for Energy Dispatch}
\label{sec:3-2}
TD3 is an off-policy deep reinforcement learning algorithm for continuous control~\cite{fujimoto2018addressing}. It uses a deterministic actor to generate actions and two critic networks to estimate their expected discounted returns, while employing a target actor and two target critics to compute the critic training targets. Here, the ED state $s_t^{ed}$ contains the current net demand $P_t^{nd}$, purchase and sale prices $p_t^p,p_t^s$, carbon intensity $c_t$, battery SOC $SOC_t^{bt}$, aggregate remaining EV energy $\sum_{i\in\mathcal S_t^{ev}}E_{i,t}^{ev}$, and two time features representing the time of day and day of week. Following the history-based state representation in~\cite{yao2026hierarchical}, these eight current features are supplemented with net demand, purchase and sale prices, and carbon intensity from six historical time steps.

The actor network $\mu_\phi$ outputs $a_t^{ed}\in[-1,1]$, where positive values indicate discharge and negative values indicate charge. The battery powers are obtained via
\begin{equation}
\small
P_t^{bt,ch}=[-a_t^{ed}]_+P^{bt,max},\quad
P_t^{bt,dc}=[a_t^{ed}]_+P^{bt,max}.
\label{eq:17}
\end{equation} 
The mapped powers satisfy \eqref{eq:5} and are further clipped, when necessary, to satisfy the SOC-dependent limits in \eqref{eq:8}. The resulting battery powers then determine grid exchange through \eqref{eq:10} and the SOC update through \eqref{eq:7}.

The reward measures financial and carbon benefits of the dispatch action relative to a baseline with an idle battery \cite{yao2025unified}
\begin{equation}
\small
B_t^{gd}=C_t^{gd,base}-C_t^{gd},
B_t^{ce}=C_t^{ce,base}-C_t^{ce},
\label{eq:18}
\end{equation}
\begin{equation}
\small
r_t^{ed}=B_t^{gd}+\omega^{ce}B_t^{ce},
\label{eq:19}
\end{equation}
where $B_t^{gd}$ is the electricity-cost saving, $B_t^{ce}$ is the reduction in net grid-related emissions, and $\omega^{ce}\geq0$ weights the carbon benefit relative to the financial benefit. The baseline uses the same EV demand and PV generation with $P_t^{bt,ch}=P_t^{bt,dc}=0$, so $P_t^{gd}=P_t^{nd}$. Surplus PV is exported in the baseline as well. The actual and baseline electricity costs are calculated using \eqref{eq:11}, and the corresponding net emissions using \eqref{eq:12}.

During training, transitions $(s_t^{ed},a_t^{ed},r_t^{ed},s_{t+1}^{ed})$ are stored in the off-policy replay buffer shown in Fig.~\ref{fig:2}. At interval $t+1$, SFGH determines the new charging demand, which forms part of $s_{t+1}^{ed}$. Mini-batches sampled from the buffer are used to train the actor $\mu_\phi$ and twin critics $Q_{\theta_1},Q_{\theta_2}$. TD3 first computes a smoothed target action
\begin{equation}
\small
\widetilde a_{t+1}^{ed}
=\operatorname{clip}_{[-1,1]}
\bigl(\mu_{\phi'}(s_{t+1}^{ed})+\nu_t\bigr),
\label{eq:20}
\end{equation}
and the target value used to train both critics is then
\begin{equation}
\small
y_t=r_t^{ed}+\gamma(1-d_t)
\min_{j=1,2}
Q_{\theta_j'}(s_{t+1}^{ed},\widetilde a_{t+1}^{ed}),
\label{eq:21}
\end{equation}
where $\mu_{\phi'}$ denotes the target actor network and
$Q_{\theta_j'}$ denotes the $j$-th target critic network. Here, $\nu_t$ is clipped Gaussian noise for target-policy smoothing, and $\gamma$ is the discount factor. The indicator $d_t$ equals one when the transition ends the simulation episode and zero otherwise. Thus, at a terminal transition, $y_t=r_t^{ed}$ without a future-return term.

Each critic minimizes the squared difference between its estimate $Q_{\theta_j}(s_t^{ed},a_t^{ed})$ and $y_t$. The actor is updated less frequently than the critics to maximize $Q_{\theta_1}(s^{ed},\mu_\phi(s^{ed}))$. After each actor update, the target-network parameters are softly updated toward the corresponding actor and critic parameters~\cite{fujimoto2018addressing}. Exploration noise is added to actor outputs during training but omitted during evaluation.

\section{Simulation Studies}
\label{sec:4}
\subsection{Simulation Settings} \label{sec:4-1}
Simulations are conducted at a resolution of $\Delta t=0.25$~h using South Wales case-study data constructed following~\cite{yao2026hierarchical}. The datasets combine real-world electricity purchase prices with synthetic transport-demand profiles. Vehicle arrivals are generated by non-homogeneous Poisson processes, while individual EV energy requirements are sampled from stochastic demand models. PV output is calculated by applying the PV generation model to real-world meteorological data. This procedure produces one year-long training set and three year-long test sets, denoted Test~1, Test~2, and Test~3~\cite{yao2026hierarchical}. For each dataset, electricity sale prices are simulated as $p_t^s=0.5p_t^p(1+\epsilon_t)$ following~\cite{yao2025unified}, where $\epsilon_t$ is a zero-mean Gaussian AR(1) process with lag-one correlation 0.90 and marginal variance 0.20. Grid carbon intensity $c_t$ is obtained from the NESO Carbon Intensity API~\cite{neso2026carbon}.

Following the station configuration in~\cite{yao2026hierarchical}, the charging facilities have $N^{ec}=5$, $P^{ev,rt}=90$~kW, and $\eta^{ev}=0.99$. The battery parameters are $E^{bt,cp}=900$~kWh, $P^{bt,max}=450$~kW, and $\eta^{bt,ch}=\eta^{bt,dc}=0.95$, with $SOC^{bt,min}=0.10$ and $SOC^{bt,max}=0.95$. Here, the scheduling and dispatch weights are set to $\kappa=1$ and $\omega^{ce}=0.5$, respectively. TD3 actor and critic networks use two hidden layers of 256 units. Training uses a learning rate of $3\times10^{-4}$, a batch size of 256, $\gamma=0.995$. The actor is updated every two critic updates, and the soft-update coefficient is 0.005. Exploration starts with 2,000 random actions, followed by Gaussian noise decreasing from 0.15 to 0.03. Each TD3 policy is trained for 15000 epochs, each containing up to 768 decision intervals. The replay buffer holds up to 100,000 transitions. All experiments are conducted on a workstation equipped with an Intel Core i7-14700F CPU, 32~GB RAM, and an NVIDIA RTX 5070 Ti GPU.

\subsection{Simulation Evaluation}\label{sec:4-2}
Table~\ref{tab:results} compares seven VS--ED combinations. First-in--first-out (FIFO) is a conventional heuristic that serves EVs in arrival order. Rule-based (RB) ED prioritizes storing surplus PV when feasible and uses the one-third and two-thirds quantiles of electricity prices/grid carbon intensity as charging and discharging thresholds, respectively. No ESS disables battery operation. PPO~\cite{schulman2017proximal} and TD3~\cite{fujimoto2018addressing} are adopted as representative on-policy and off-policy DRL algorithms, respectively. For each learned configuration, five policies trained with different random seeds are evaluated on each test set. Results are reported as the mean and standard deviation across these policies, while each deterministic baseline is evaluated once per test set. Annual performance is evaluated using the cumulative service delays $C^{svc}=\sum_t(C_t^{wt}+C_t^{ot})$, net electricity cost $C^{gd}=\sum_t C_t^{gd}$, and net grid-related carbon emissions $C^{ce}=\sum_t C_t^{ce}$, where the summations cover all intervals in each year-long test set.
\begin{table*}[!t]
\caption{Annual outcomes in each test set: mean $\pm$ sample standard deviation over five runs}
\label{tab:results}
\centering
\footnotesize
\setlength{\tabcolsep}{2.7pt}
\renewcommand{\arraystretch}{1.05}
\begin{tabular}{@{}lrrrrrrrrr@{}}
\toprule
& \multicolumn{3}{c}{Test 1}
& \multicolumn{3}{c}{Test 2}
& \multicolumn{3}{c}{Test 3}\\
\cmidrule(lr){2-4}
\cmidrule(lr){5-7}
\cmidrule(lr){8-10}
Approach
& $C^{svc}$ (h) & $C^{gd}$ (kGBP)& $C^{ce}$ (tCO$_2$)
& $C^{svc}$ (h)& $C^{gd}$ (kGBP)& $C^{ce}$ (tCO$_2$)
& $C^{svc}$ (h)& $C^{gd}$ (kGBP)& $C^{ce}$ (tCO$_2$)\\
\midrule







SFGH--TD3
& $\mathbf{4379\!\pm\!0}$ & $\mathbf{214.81\!\pm\!0.21}$ & $\mathbf{365.21\!\pm\!0.12}$
& $\mathbf{3737\!\pm\!0}$ & $\mathbf{211.87\!\pm\!0.21}$ & $\mathbf{361.82\!\pm\!0.25}$
& $\mathbf{3890\!\pm\!0}$ & $\mathbf{212.49\!\pm\!0.16}$ & $\mathbf{362.33\!\pm\!0.23}$\\

FIFO--RB
& $4497$ & $222.34$ & $372.78$
& $3848$ & $219.15$ & $369.51$
& $4004$ & $220.48$ & $370.44$\\

SFGH--No ESS
& $\mathbf{4379}$ & $234.79$ & $387.17$
& $\mathbf{3737}$ & $231.69$ & $383.92$
& $\mathbf{3890}$ & $233.05$ & $384.62$\\

SFGH--RB
& $\mathbf{4379}$ & $222.34$ & $372.78$
& $\mathbf{3737}$ & $219.15$ & $369.52$
& $\mathbf{3890}$ & $220.50$ & $370.44$\\

PPO--PPO
& $4423\!\pm\!40$ & $216.25\!\pm\!1.39$ & $369.33\!\pm\!0.67$
& $3782\!\pm\!32$ & $212.57\!\pm\!1.41$ & $366.01\!\pm\!0.71$
& $3928\!\pm\!27$ & $214.26\!\pm\!1.41$ & $366.88\!\pm\!0.59$\\

PPO--TD3
& $4566\!\pm\!6$ & $222.12\!\pm\!0.29$ & $365.42\!\pm\!0.23$
& $3969\!\pm\!7$ & $218.81\!\pm\!0.22$ & $362.41\!\pm\!0.27$
& $4104\!\pm\!6$ & $220.79\!\pm\!0.12$ & $363.05\!\pm\!0.26$\\

TD3--TD3
& $4573\!\pm\!15$ & $215.10\!\pm\!0.19$ & $365.89\!\pm\!0.27$
& $3959\!\pm\!11$ & $212.07\!\pm\!0.38$ & $362.43\!\pm\!0.16$
& $4089\!\pm\!9$ & $213.19\!\pm\!0.32$ & $363.43\!\pm\!0.37$\\
\bottomrule
\multicolumn{8}{l}{\scriptsize
Discrepancies of up to 1 may occur due to rounding.}

\end{tabular}
\end{table*}

On the VS side, SFGH achieves the lowest $C^{svc}$ in all three test sets, reducing service delays by 2.62\%, 2.88\%, and 2.85\% relative to conventional FIFO. With ED fixed to TD3, SFGH--TD3 improves on PPO--TD3 by
4.10\%, 5.85\%, and 5.21\%, and on TD3--TD3 by
4.24\%, 5.61\%, and 4.87\%, respectively. It also reduces service delays by 0.99\%, 1.19\%, and 0.97\% compared with PPO--PPO. SFGH's advantage over PPO and TD3 may arise from directly evaluating waiting and overflow reductions, which suits the scheduling task considered in the EV charging station.

On the ED side, SFGH--TD3 achieves the lowest mean net cost and net emissions in all test sets. With VS fixed to SFGH, TD3 reduces cost and emissions by 8.51--8.82\% and 5.67--5.80\%, respectively, compared with operation without storage (No ESS). The corresponding reductions relative to SFGH--RB are 3.32--3.63\% and 2.03--2.19\%. SFGH--TD3 also outperforms multi-agent DRLs (PPO--TD3, PPO--PPO, and TD3--TD3) on both ED metrics. One possible explanation concerns the demand distribution supplied by VS: if the VS policy continues to learn over time, changes in the resulting demand profiles may complicate ED learning. A fixed SFGH rule avoids this additional source of variation. Together, these results support combining SFGH for VS with TD3 for ED to improve service, financial, and carbon outcomes.

\begin{figure}[]
\centering
\includegraphics[width=9.0cm, trim=2 2 2 2, clip]{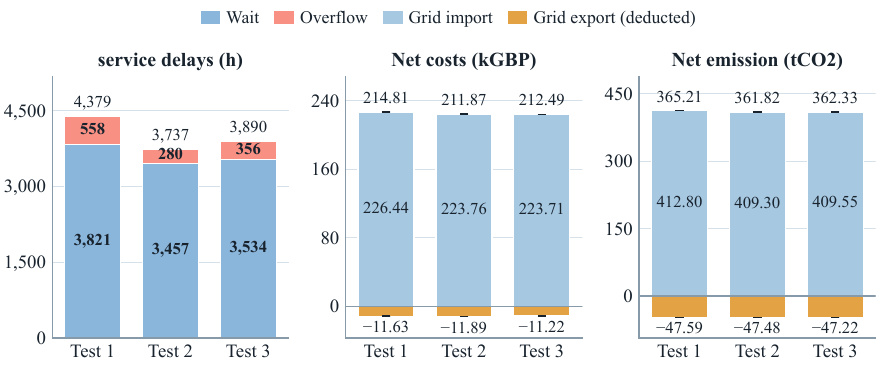}
\caption{Decomposition of annual SFGH--TD3 outcomes}
\label{fig:3}
\end{figure}
Fig.~\ref{fig:3} further decomposes the annual outcomes of SFGH--TD3. In the left subfigure, waiting accounts for most of the service delays because it accumulates throughout each EV's time in the queue, whereas overflow accrues only beyond $w^{max}$. The middle subfigure shows that export revenue offsets 11.22--11.89~kGBP of annual grid purchase costs, bringing net costs to 211.87--214.81~kGBP. The right subfigure shows export carbon credits of 47.22--47.59~tCO$_2$, yielding net emissions of 361.82--365.21~tCO$_2$ under the adopted accounting convention. These components highlight how bidirectional grid exchange contributes to both the financial and carbon objectives of the charging station.

\subsection{Operating Performance}\label{sec:4-3}
\begin{figure}[]
\centering
\includegraphics[width=9.0cm, trim=7 10 10 7, clip]{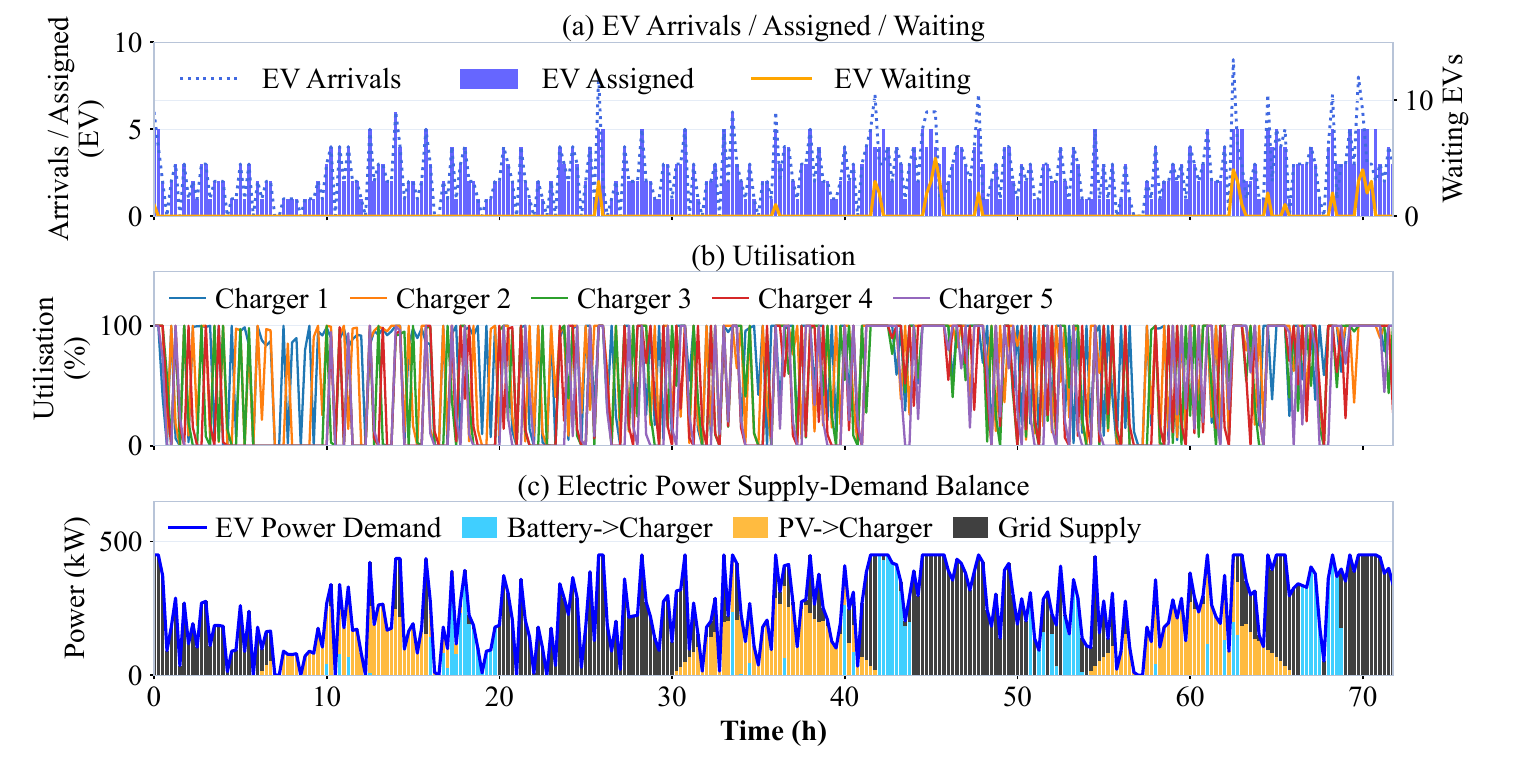}
\caption{Vehicle scheduling and charging power balance}
\label{fig:4}
\end{figure}
\begin{figure}[]
\centering
\includegraphics[width=9.0cm, trim=7 10 10 7, clip]{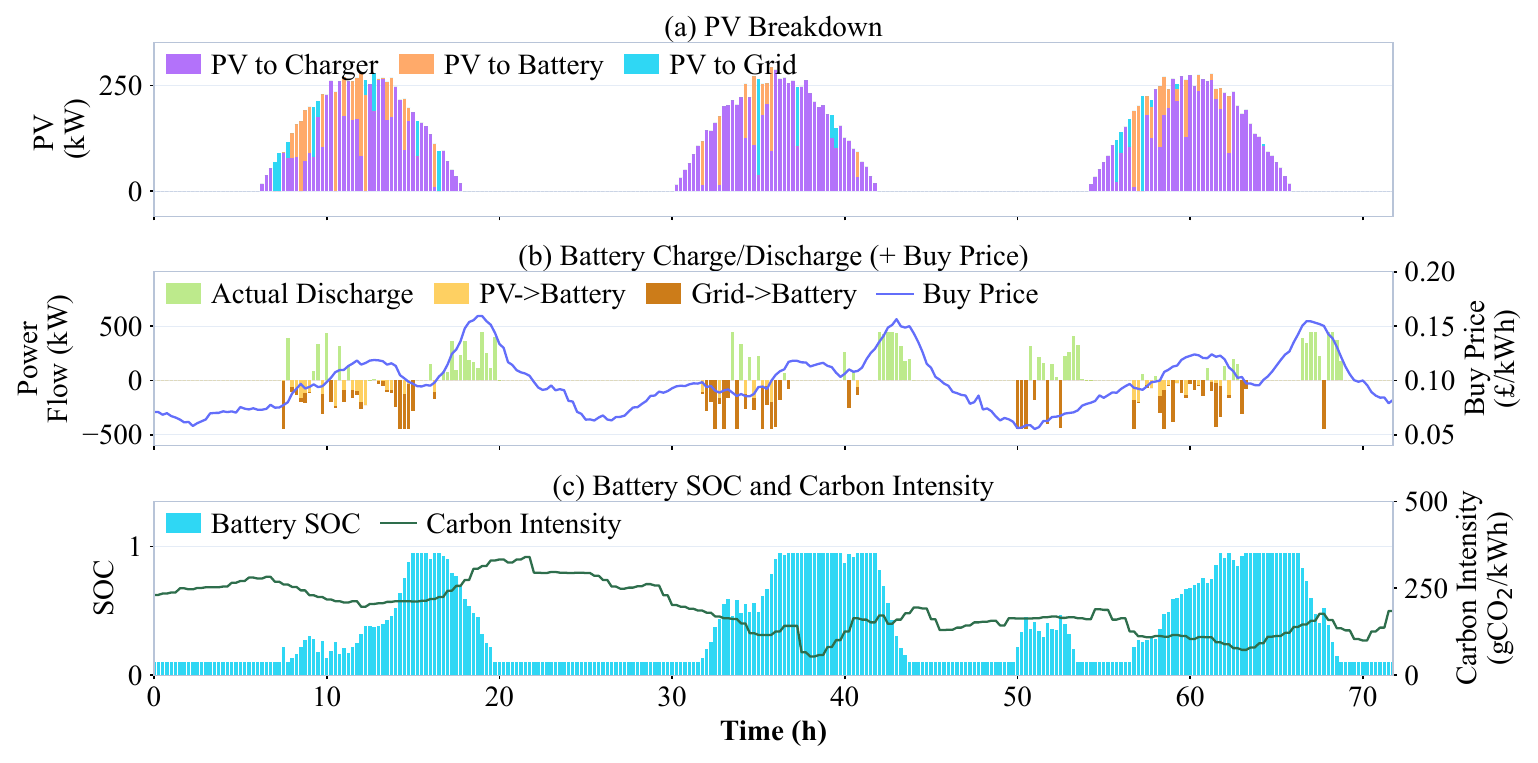}
\caption{PV allocation and battery operation}
\label{fig:5}
\end{figure}
Figs.~\ref{fig:4} and~\ref{fig:5} illustrate SFGH--TD3 operation over a representative 72-h period. In Fig.~\ref{fig:4}(a), queues formed during busy periods clear as chargers become available in Fig.~\ref{fig:4}(b), indicating that SFGH handles arrival bursts without a persistent backlog during this period. Fig.~\ref{fig:4}(c) shows that PV generation and battery discharge partially supply EV charging demand, reducing reliance on direct grid purchases.

Fig.~\ref{fig:5}(a) shows PV primarily supplying EV charging, with surplus stored or exported to accommodate mismatches between PV generation and demand. Charging before purchase-price peaks and discharging near those peaks in Fig.~\ref{fig:5}(b) illustrates the mechanism for reducing net cost through energy arbitrage. Charging at relatively low carbon intensity and then discharging as it rises in Fig.~\ref{fig:5}(c) provides an additional opportunity to reduce net emissions. The SOC plateaus also show that storage capacity limits how much energy can be shifted across intervals. These operating patterns confirm the effectiveness of SFGH--TD3 in managing vehicle scheduling and energy dispatch.

\section{Conclusion}
This paper presented an SFGH--TD3 approach for vehicle scheduling and carbon-aware energy dispatch at a PV--battery EV charging station with bidirectional grid exchange. SFGH determines charger assignments based on waiting and overflow reductions, and TD3 controls battery operation within feasible limits using electricity prices and carbon intensity. Various simulation studies support the effectiveness of the proposed SFGH--TD3 approach. Future work will incorporate battery degradation and grid capacity constraints to assess their effects on scheduling and dispatch performance.

\begingroup
\interlinepenalty=10000
\bibliographystyle{IEEEtran}
\bibliography{references}
\endgroup
\end{document}